\documentclass[prl,twocolumn]{revtex4} \usepackage{graphicx}
\begin{document} \title{How strongly correlated is
MnSi?} \author{F. Carbone$^1$, M. Zangrando$^2$, A. Brinkman$^1$, A.
Nicolaou$^2$, F. Bondino$^2$, E. Magnano$^2$, A.A. Nugroho$^{4,5}$,
F. Parmigiani$^{2,3}$, Th. Jarlborg$^1$, D. van der Marel$^1$}
\affiliation{$^1$D\'epartement de Physique de la Mati\`ere
Condens\'ee, Universit\'ee de Gen\`eve, CH-1211 Gen\`eve 4,
Switzerland\\$^2$Laboratorio Nazionale TASC-CNR, Basovizza S.S. 14,
Km 163.5, 34012 Trieste, Italy\\$^3$INFM, Dipartimento di Matematica
e Fisica, UCSC, Via dei Musei 41, 25121 Brescia,
Italy\\$^4$Materials Science Centre, University of Groningen, 9747
AG Groningen, The Netherlands\\$^5$Jurusan Fisika, Institut
Teknologi Bandung, Indonesia} \date{September 9, 2005}
%

\begin{abstract} We present an experimental study of the
electronic structure of MnSi. Using X-ray Absorption Spectroscopy,
X-ray photoemission and X-ray fluorescence we provide experimental
evidence that MnSi has a mixed valence ground state. We show that
self consistent LDA supercell calculations cannot replicate the
XAS spectra of MnSi, while a good match is achieved within the
atomic multiplet theory assuming a mixed valence ground state. We
discuss the role of the electron-electron interactions in this
compound and estimate that the valence fluctuations are suppressed
by a factor of 2.5, which means that the Coulomb repulsion is not
negligible.

\end{abstract}
\maketitle

%
%
Traditionally the magnetism of MnSi is considered as weakly
itinerant\cite{moriya73,taillefer86}, {\em i.e.} the
spin-polarization is modeled as a relative shift of bands of
delocalized Bloch-states for the two spin-directions. At ambient
pressure MnSi orders heli-magnetically below $T_C$=29.5 K, and
becomes ferromagnetic in a magnetic field exceeding 0.6 Tesla. The
Hall effect and the negative magneto-resistance\cite{manyala00} in
the ferromagnetic phase agree well with the theory of
spin-fluctuations in itinerant ferromagnetism\cite{moriya73}. Also
the inelastic neutron scattering data can be interpreted in this
framework\cite{ishikawa85}. The saturation moment of the
magnetically ordered phase is 0.4 $\mu_B$ per Mn atom. On the other
hand, {\em ab initio} calculations based on the the Local Density
Approximation (LDA) indicate a tendency of the Mn-atoms to form a
moment close to 1 $\mu_B$ if the real lattice constant for MnSi
(4.558 $\AA$) is used\cite{jeong04,lerch94}. A fit of the
susceptibility in the paramagnetic phase to a Curie-Weiss law gives
2.2 $\mu_B$ per Mn atom\cite{wernick72}.

Recently, several properties of MnSi have been discovered which had
not been anticipated on the basis of the itinerant model and which
remain to be fully understood: Above 14.6 kbar the material enters a
phase with partial heli-magnetic order along the (1,1,0)
direction\cite{pfleiderer04}, where the electrical resistivity is
proportional to T$^{3/2}$ in contradiction to standard notions of a
Landau Fermi liquid\cite{doiron03}. A further indication of
anomalous low energy scale properties follows from the non-Drude
infrared optical conductivity at ambient pressure\cite{mena03},
proportional to $(i\omega)^{-1/2}$. Above $T_C$ the resistivity is
described by the formula\cite{mena03} $\rho= \rho_{sat}T/(T_0+T)$
which for $T \gg T_0=180K$ approaches the Mott-Ioffe-Regel limit,
$\rho_{sat}$ =287$\mu\Omega$cm. The rapid rise towards saturation
corresponds to a strong dissipation of the charge transport. The
abrupt drop of the electrical resistivity when the material is
cooled through $T_C$ suggests that this dissipation is due to a
coupling to magnetic fluctuations.

Here, using X-ray Absorption Spectroscopy (XAS), X-ray photoelectron
spectroscopy (XPS) and X-ray fluorescence spectroscopy (XFS), we
provide the experimental evidence that MnSi has a mixed valence
ground state which cannot be described by the standard LDA approach.
We will show that the electron-electron correlations are not
negligible, the value of U/W is estimated to be around 0.4, where U
are the on site Coulomb repulsion and W is the bandwidth; we think
that most likely the observed deviations from the usual itinerant
picture are due to the suppression of valence fluctuations in the
ground state of the material.

\begin{figure}[ht]
   \centerline{\includegraphics[width=10cm,clip=true]{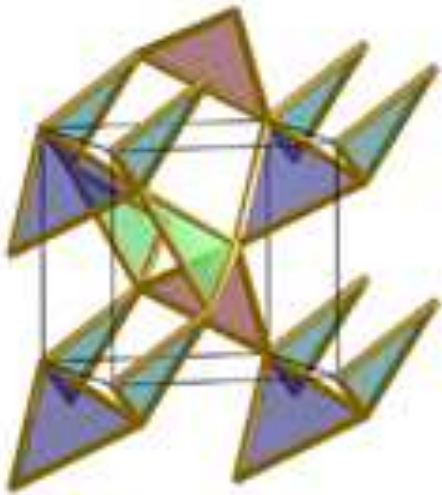}}
   \caption{(Color). Mn sublattice of MnSi. The corners of the triangles, all of which are equilateral, correspond to the
   positions of the Mn-atoms.}
  \label{rendered}
 \end{figure}

The crystal structure of MnSi is generated by the cubic $B20$
structure \cite{boren33,vandermarel98}. The unit cell contains 4
Mn atoms at crystallographically equivalent positions. The
sub-lattice of transition metal atoms, displayed in Fig.
\ref{rendered}, reveals that the basic structural element is an
equilateral triangle of 3 Mn atoms. The structure is
corner-sharing: Each Mn-atom connects 3 triangles, which occur
with 4 different orientations along the body-diagonals of the
cubic unit cell. The singly connected loops of the structure shown
in Fig. \ref{rendered} contain an odd number of bonds. The
structural similarity to the
pyrochlore\cite{obradors88,ramirez90},
Kagome\cite{broholm90,chalker92}, Gadolinium Gallium
Garnet\cite{schiffer95}, and the $\beta$-Mn
lattices\cite{nakamura97,canals00} is a peculiarity that has been
overlooked so far. This might play a role in the formation of the
helical magnetic structure observed below 29 K.

%
MnSi high quality single crystals were grown by the floating zone
technique starting from 4N purity Mn and 5N purity Si. All samples
were characterized by x-ray diffraction, EDX elemental analysis and
electrical resistivity. The residual resistivity of all MnSi samples
was less than 2 $\mu\Omega cm$.

The experiments were performed at the BACH beam line
\cite{Zangrando1} of the ELETTRA synchrotron in Trieste. XAS  was
performed in total electron yield (TEY), measuring roughly the first
50 {\AA} of the surface, and total fluorescence yield (TFY),
measuring down to 200 nm in the bulk. The XAS spectra were
normalized to the incident photon flux, the resolution in TEY was
150 meV and 400 meV in TFY. The fluorescence experiments were done
recording the fluorescent decay of Mn $3d\longrightarrow 2p$ and
$2p\longrightarrow 3s$ levels on a CCD detector.


\begin{figure}[ht]
   \centerline{\includegraphics[width=9cm,clip=true]{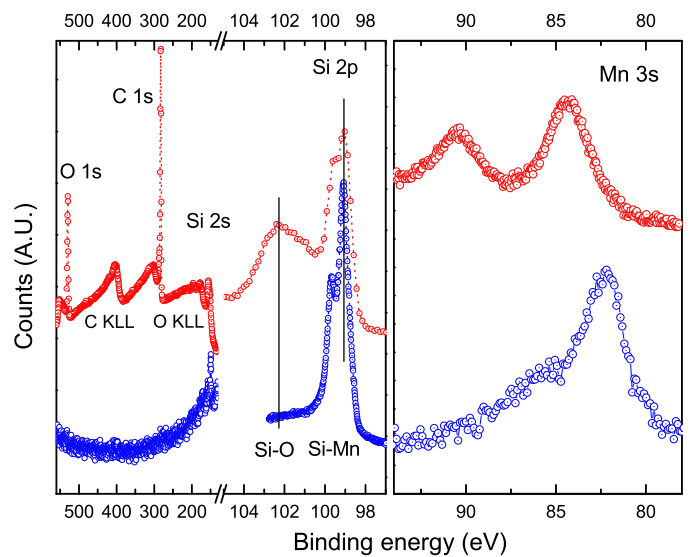}}
   \caption{(Color online). XPS spectra of MnSi before and after cleaving. In the right part of the figure
   one can see the high resolution spectra of the Mn 3$s$ levels measured with an incident photon energy of 418 eV and the Si 2$p$ levels  measured with an incident photon energy of 196 eV
   before cleaving and 142 eV after cleaving; in the left part a survey from the Si 2$s$ to the O 1$s$ is displayed measured at 655 eV incident photon energy. The blue curve
   represents the spectrum after cleaving, the red curve was recorded before cleaving. After cleaving
   the high binding energy component of the Si 2p line is suppressed , the Mn 3s level splitting diminishes and the C and O 1$s$ lines
   are also suppressed.}
\label{xps_surf}
\end{figure}

Large single crystals were cleaved in situ prior to the measurements
in order to obtain clean surfaces; the surface quality was checked
with XPS, Fig. \ref{xps_surf}. The base pressure in the measurement
chamber was $1\cdot10^{-10}$ mbar. XAS and XPS spectra were recorded
at room temperature within minutes after cleaving. The contamination
of the surface before and after cleaving was checked by oxygen and
carbon 1$s$ photoemission. The cleaved surface of the sample was
scanned spatially with steps of 100 $\mu m$ and XPS was recorded at
each position. This analysis showed that a significant carbon
contamination is present on the border of the sample. This
contamination affects dramatically the shape of the TEY XAS. Only at
least 150 $\mu m$ away from the sample's border, where the XPS
reveals a very clean surface, we could have a TEY spectrum in
agreement with the TFY one, representative of the bulk properties of
the material. In the XPS spectra, recorded in the middle of a
cleaved sample, the oxygen and carbon $1s$ lines are completely
suppressed with respect to the non cleaved sample, as shown in Fig.
\ref{xps_surf}. The analysis of the surface revealed that carbon,
MnO and SiO$_2$ are the main contaminants. The XPS of Si 2p levels
shows a component around 102 eV associated to SiO$_2$; the Mn 3s
splitting on the sample before cleaving was 6.3 eV, in agreement
with earlier reports for MnO \cite{galakov02}. In the cleaved sample
the high binding energy peak of the Si 2p level is suppressed, the
Mn 3s levels splitting diminishes to a much smaller value and the
carbon and oxygen 1$s$ lines are suppressed.


\begin{figure*}[tb]
   \centerline{\includegraphics[width=17cm,clip=true]{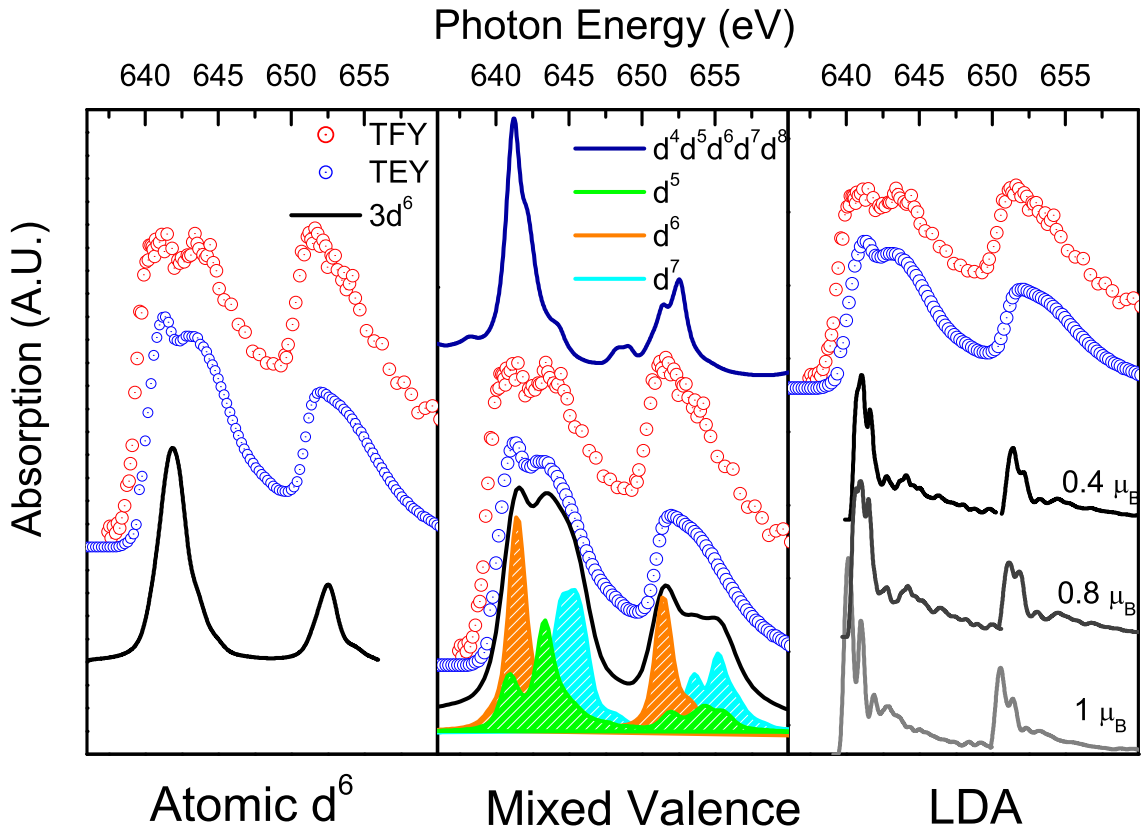}}
     \caption{(Color). Left panel: Mn $L_{2,3}$ edge measured XAS together with atomic multiplet calculation for a $3d^6$ ground state.
     The TFY experiment has a resolution of 0.4 eV (red open symbols); the TEY experiment has a resolution of 0.2 eV (blue open symbols).
     Middle panel: the experimental spectra are plotted together with the Mn mixed valence atomic multiplets calculations in a cubic crystal
     field (black line); below this line is possible to see the contribution from the different configurations. The dark blue line represents the
     superposition of the $d^4, d^5, d^6$ and $d^7$ configurations with the weights given by the binomial distribution in table \ref{lda_exp}, which correspond
     to the non interacting particle picture.
     Right panel: the LDA calculations are plotted for 3 different value of the lattice parameter together with the experimental spectra.}
\label{xasmn}
 \end{figure*}

\begin{figure}[ht]
   \centerline{\includegraphics[width=9cm,clip=true]{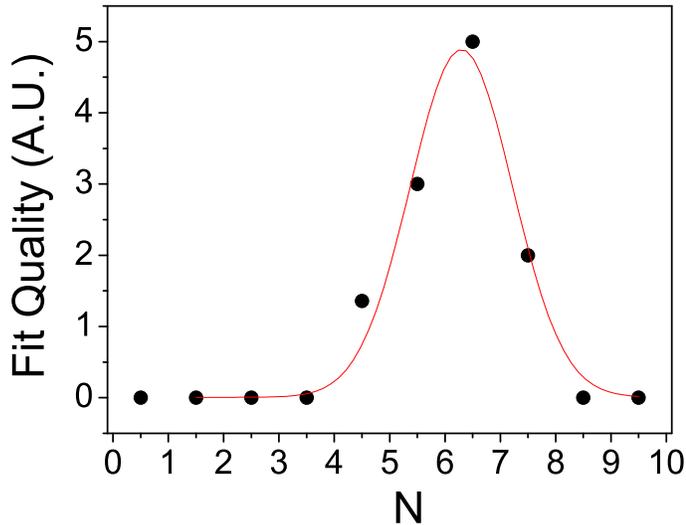}}
     \caption{(Color online). We display the inverse $\chi^2$ for the fits to the
     experimental data of the superposition of ${d^4+d^5}$, ${d^5+d^6}$, ${d^6+d^7}$ and ${d^7+d^8}$ respectively.}
\label{d4d8}
 \end{figure}

In Fig. \ref{xasmn} we display the Mn $L_{2,3}$ XAS spectrum of MnSi
measured both in TEY and TFY; one can see that the two spectra are
almost identical, indicating that we are probing indeed the bulk.
The two main peaks correspond to the $2p_{1/2}$ (642 eV) and
$2p_{3/2}$ (653 eV) spin-orbit split components of the 2p core
level. In a one particle picture these two edges have the same
spectral shape, as illustrated by a first principles calculation
using the Local Density Approximation (LDA, black line in Fig.
\ref{xasmn}). Self consistent LDA-LMTO (Local Density Approximation
- Linear Muffin Tin Orbital) calculations have been performed for
64-atom supercells; one of the Mn atoms has a core hole. The
groundstate of the calculation was ferromagnetic, adopting three
different states of magnetic polarization characterized by local
moments of 0.4, 0.8 and 1 $\mu_B$, labelled as such in Fig.
\ref{xasmn}. The XAS spectrum corresponds to a broadened sum of the
unoccupied local spin Mn-d DOS functions. A known problem of band
calculations in MnSi is the predicted value of the local moment on
the transition metal atom \cite{lerch94}. This quantity is strongly
dependent on the unit cell dimension and tends to be higher then the
measured one when the lattice constant has the experimentally
determined dimension of 4.558 \AA. We checked the influence of this
effect on the XAS spectrum in 3 cases, changing the lattice
constant: the local moment of Mn is 0.4 $\mu_B$ (the experimentally
measured value) for a lattice parameter $a=4.36 \AA$, 0.8 for $a=4.5
\AA$ and 1 $\mu_B$ for the measured lattice constant $a=4.55 \AA$.
This is shown in the right panel of Fig. \ref{xasmn}; this effect
weakly modifies the XAS spectrum and cannot explain the strong
departure from the measured one. It is evident that LDA calculations
are narrower and cannot replicate the XAS spectra for MnSi. In the
middle panel of Fig. \ref{xasmn} we compare the experimental spectra
with atomic model calculations performed with a standard computer
program \cite{cowan81}. We calculate the XAS spectra for several
different configurations: Mn $3d^4, 3d^5, 3d^6, 3d^7$ and $d^8$ in a
cubic crystal field environment of 2.4, 2.6 and 3 eV. Furthermore
least mean square fits to the data of the weighted superposition of
4 single valence spectra, $d^4, d^5, d^6, d^7$ and $d^5, d^6, d^7,
d^8$ were performed. The least mean square routine tends to give a
negligible weight to the $d^4$ and $d^8$ configurations. We estimate
the error bars of this approach as the maximum spread of values
obtained for the $d^5, d^6, d^7$ configurations in the two cases for
the 3 mentioned values of the crystal field. The crystal field is
estimated from the band splitting observed in the high symmetry
points of the band calculations \cite{jeong04}. In the best fit the
relative weights of the different valences are found to be: $0\%
d^4, 21\% d^5, 55\% d^6, 24\% d^7$, $0\%d^8$ in a crystal field of
2.6 eV. In Fig. \ref{d4d8} we plot the inverse of the $\chi^2$
obtained fitting the experimental data to the combination of
${d^4+d^5}$, ${d^5+d^6}$, ${d^6+d^7}$ and ${d^7+d^8}$ respectively.
This calculation shows that the fitting quality is peaked around the
$d^6$ configuration and supports the conclusion that a large
contribution to the XAS spectrum comes from the $3d^6$
configuration.

\begin{figure}[ht]
   \centerline{\includegraphics[width=9cm,clip=true]{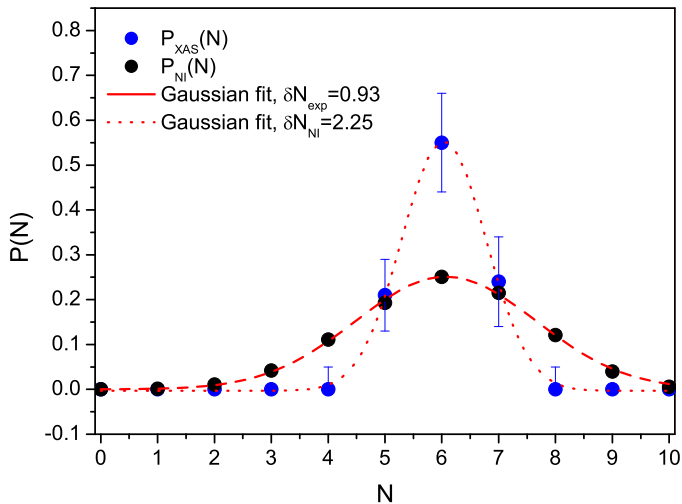}}
     \caption{(Color online). The theoretical and experimentally derived values of
     P(N) are plotted together with the fit to Eq. 1. From these fits
     we extract the values for $\delta N$ and thus $\frac{\delta N_{NI}}{\delta N_{exp}}$ = 2.6}
\label{ldaxas}
 \end{figure}

In the left panel of Fig. \ref{xasmn} we display a calculation for
an atomic $3d^6$ ground state; this simple calculation also does
not represent satisfactorily the experiments. The better agreement
between the experiments and the atomic multiplet mixed valence
calculation emphasizes two important properties of the electronic
configuration of MnSi: (i) the dominant configuration is $3d^6$;
(ii) experimentally, the valence fluctuations are given by:
\begin{equation} p(N) = P(N_0)\cdot exp[-(\frac{N-N_0}{\delta
N})^2] \end{equation} where $\delta N_{exp}=0.92$ and $N_0=6$. For
non interacting particles distributed over 10 $3d$ bands, having
the average occupation of 6 electrons ($N_0=6$), P(N) is given by
the binomial equation: \begin{equation} P_{NI}(N) =
0.6^N0.4^{10-N}10!\frac{N!}{N!(10-N)!} \end{equation} (NI = non
interacting), which is to a very good approximation given by Eq. 1
with $N_0=6$ and $\delta N_{NI} = 2.3$. Thus the value
$\frac{\delta N_{NI}}{\delta N_{exp}}=2.5$ gives a measure of the
valence suppression in the ground state. In table \ref{lda_exp} we
show the probability of having N electrons on an ion as a function
of the occupation number in a LDA picture, together with the
experimental findings. In Fig. \ref{ldaxas} one can see the fit to
Eq. 1 for the experimentally derived P(N) and the theoretical
ones. The sharp suppression of valence fluctuations in the ground
state of Mn observed experimentally is likely the consequence of
the on-site Coulomb interaction in the 3d shell of Mn. For the
$d^6$ configuration of Mn $U_{eff} = F_0-J-C = 1 eV$ \cite{dirk},
where $F_0$ is the intra-shell Coulomb repulsion, J is the
intra-shell exchange interaction and C takes into account all the
multipole contributions of the Coulomb and exchange interaction.
The overall 3d band-width of MnSi is about 6 eV, but this value in
part reflects a relative shift of the different group of bands,
representing the crystal field splitting. The width of each of the
sub-bands is approximatively 2.5 eV, hence U = 0.4W in this
compound. This value implies that MnSi has to be considered as an
itinerant system. On the other hand the valence fluctuations
should be strongly suppressed as compared to the non interacting
picture, and this indeed corresponds to what we observed
experimentally. As a result of final state interference effects
\cite{zaanen,kotani}, the values of P(5) and P(7) given in table
\ref{lda_exp} and Fig. 3 are probably somewhat underestimated.

\begin{table} \caption{Theoretical P(N) assuming
non interacting particles, [$P_{NI}(N)$], experimental P(N)
obtained from the mixed-valence fit to the XAS spectrum,
$[P_{exp}(N)]$. The values of $\Delta(N)$ correspond to the shift
of the energies $E(2p\longrightarrow 3d^{N+1})$, with respect to
the output of the Cowan code, of the final state multiplets; the
cubic crystal field parameter was 2.6 eV for all configurations.}
\begin{tabular}{lccccccccccc}
  \hline\hline
       & N & $P_{NI}(N)$ & $P_{exp}(N)$ & $\Delta(N)$ &\\
  \hline
  & 0 & 0.0001 & - & - &\\
  & 1 & 0.0015 & - & - &\\
  & 2 & 0.011 & - & - &\\
  & 3 & 0.042 & - & - &\\
  & 4 & 0.111 & - & - &\\
  & 5 & 0.193 & 0.21 & 2 &\\
  & 6 & 0.251 & 0.55 & 0.38 eV &\\
  & 7 & 0.215 & 0.24 & 3.72 eV &\\
  & 8 & 0.121 & - & - &\\
  & 9 & 0.04 & - & - &\\
  & 10 & 0.006 & - & - &\\
  \hline\hline
\end{tabular}
\label{lda_exp}
\end{table}

\begin{figure}[ht]
   \centerline{\includegraphics[width=9cm,clip=true]{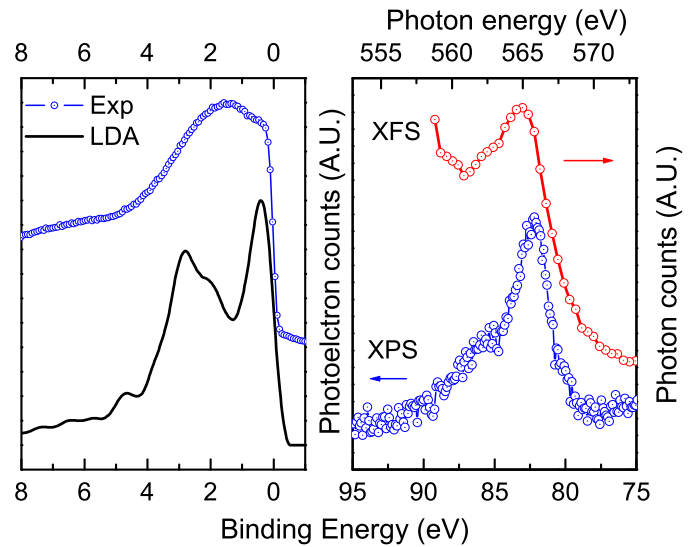}}
     \caption{(color online). Top panel: Valence Band photoemission measured at an incident photon energy of 104 eV together with LDA calculations. Lower panel: Photoemission
     (blue open symbols) measured at 418 eV incident photon energy and fluorescent (red open symbols) spectra of Mn 3s levels measured at 660 eV incident photon energy.}
\label{XPSn}
\end{figure}

In Fig. \ref{XPSn} we present the photoemission spectrum of the Mn
3s core level measured at an incident photon energy of 418 eV  and
the fluorescence spectrum measured at a photon energy of 660 eV;
since photoemission is a very surface sensitive technique, we cross
check our results acquiring the corresponding fluorescence spectrum
when possible. The Mn 3s photoemission shows a shoulder on the high
energy side of the spectrum. Most likely, the mixed valence ground
state we discussed before is responsible for this weak shoulder
visible in the 3s spectrum. The asymmetry of the 3s levels
photoemission in insulating Mn compounds, such as MnO, MnF$_2$ or
manganites, has been shown to be caused by the many-body interaction
between the core-hole and the localized 3d electrons
\cite{galakov02,parmigiani}. In this case the role of the exchange
interaction is predominant and, when the orbital moment does not
contribute to the total magnetic moment of the charge carriers, a
direct relation between the 3s level splitting and the spin magnetic
moment is valid. On the other hand, it is well known that this
relation doesn't hold any longer in more metallic systems
\cite{gey}. When the electronegativity of the ligand atom decreases,
the charge transfer satellites and the screening of the final state
become more important, as a result it is not possible any longer to
attribute the peaks in the 3s spectra to pure spin states. Usually,
in more covalent systems, the 3s levels splitting is smaller than
what one would expect in the localized scenario because of these
effects. We believe that this is the case in MnSi, whose metallic
behavior reflects the covalent nature of the Mn-Si bonding.

In Fig. \ref{XPSn} we compare the experimental valence band
photoemission spectra with the LDA calculations. The calculations
includes the radial matrix elements but ignores the k-conservation
between initial and final states. This is a reasonable approximation
in the limit of large photon energy\cite{jarlborg}. In the
calculation a peak is evident around 2.8 eV away from the Fermi
edge, a similar feature is visible in the experimental spectrum,
although its position is only 1.8 eV away from the Fermi edge. The
valence band photoemission spectra have been collected using three
incoming photon energies: 86 eV, 104 eV and 196 eV and no
appreciable changes where observed. Also in this case the agreement
between the calculation and the experiment is not satisfactorily.
The valence band photoemission on MnSi has already been reported
together with the LDA calculation in Ref. \onlinecite{giappi}. The
authors point out that the major deviations from the raw spectra and
the calculations are ascribable to the on site Coulomb repulsions,
in agreement with our conclusion.

%
Our observations evidence the fact that in this class of materials
it is not justified to neglect completely the electron electron
correlations. The discrepancy between the single particle scenario
and the experiment is corroborated by the comparison in Fig.
\ref{xasmn} (a) of the LDA-prediction of the XAS spectrum to the
experimental data. It would be tempting to attribute this
discrepancy to the fact that XAS is a high energy probe, and that
the observed spectra correspond to the final state with an extra
core-hole present. However, (i) both in the band-calculation as well
as in the atomic multiplet calculations shown in Fig.\ref{xasmn} (a)
the presence of the core-hole has been taken into account, (ii)
theoretically these spectra are expected to be a very sensitive
fingerprint of the initial state electronic configuration, (iii) the
same concerns would apply to the transition metal oxide family,
where XAS has been quite successful probes of the magnetic
properties \cite{thole85,thole91,thole92,carra93}. Moreover, also
valence band photoemission, where no core hole is present, is
inconsistent with the LDA approach. The cross check of the results
by means of different techniques, electron counting and photon
counting techniques, make us confident that we are indeed probing
the electronic structure of bulk MnSi. Our estimated value for U/W
around 0.4 classifies MnSi in a class of materials where none of the
two approximations is particularly good: completely neglecting the
electron-electron interactions or considering them as dominant. The
helical magnetic structure of MnSi has been explained in terms of
the Dyaloshinskii-Moryia interaction; the interplay between
spin-orbit coupling and exchange interaction can result in an
anisotropic exchange interaction, responsible for the helical
magnetic structure in low symmetry crystals. For this to happen the
motion of the conduction electrons must have a finite orbital
component, for example a $3d^5$($^6S$) ground state would be rather
unfavorable in this context, having a null orbital moment. Our
observations are compatible with this picture, providing an
experimental support to the microscopic model.

In conclusion, we examined the electronic structure of MnSi using
XAS, XFS and XPS. The experimental data indicate that MnSi has a
mixed valence ground state of predominantly $3d^6$ character. The
suppression of the valence fluctuations indicate that a
considerable electron-electron interaction is present in this
material; we estimate that the valence fluctuations are suppressed
by a factor of 2.5, meaning that the Coulomb repulsions are non
negligible, but insufficient to form local moments on the Mn 3d
shell.

This work was supported by the Swiss National Science Foundation
through the National Center of Competence in Research "Materials
with Novel Electronic Properties-MaNEP". The authors gratefully
acknowledge stimulating discussions with F.M.F. de Groot, F.P. Mena,
G. Aeppli, J. diTusa, A. Yaouanc, P. Dalmas de Réotier, M. Laad and
technical support from J.P. Soulié, T. Pardini and M. Zacchigna.


\begin{thebibliography}{100}
%
\bibitem{moriya73} T. Moriya, and A. Kawabata, J. Phys. Soc. Jpn.,
{\bf 34}, 639 (1973).

\bibitem{taillefer86} L. Taillefer, G.G. Lonzarich, and P Strange,
J Magn. Magn. Mater. {\bf 54-57}, 957 (1986).

\bibitem{manyala00} N. Manyala, Y. Sidis, J.F. di Tusa, G. Aeppli,
D.P. Young and Z. Fisk. Nature {\bf 404}, 581 (2000).

\bibitem{ishikawa85} Y. Ishikawa, Y. Noda, Y.J. Uemera, C.F. Majkrzak, and G. Shirane,
Phys. Rev. B {\bf 31}, 5884 (1985).

\bibitem{jeong04} T. Jeong, and W. E. Pickett,
Phys. Rev. B {\bf 70}, 075114 (2004).

\bibitem{lerch94} P. Lerch, and Th. Jarlborg,
J. Magn. Magn. Mat. {\bf 131}, 321 (1994).


\bibitem{wernick72} J.H. Wernick, G.K. Wertheim and R.C. Sherwood,
Mater. Res. Bull. {\bf 7}, 1153 (1972).

\bibitem{pfleiderer04} C. Pfleiderer, D. Reznik, L. Pintschovius, H. v. Lohneysen, M. Garst, A. Rosch,
Nature {\bf 427}, 15 (2004).


\bibitem{doiron03} N. Doiron-Leyraud {\em et al.}, Nature {\bf 425}, 595 (2003).

\bibitem{mena03} F.P. Mena,D. van der Marel, A. Damascelli, M. Faeth, A.A. Menovsky, J.A. Mydosh,
Phys. Rev. B {\bf 67}, 241101 (2003)

\bibitem{boren33} B. Bor\'en, Ark. Kemi Min. Geol. {\bf 11}A, 1
(1933).

\bibitem{vandermarel98} D. van der Marel,
A. Damascelli, K. Schulte, and A. A. Menovsky, Physica B {\bf 244},
138 (1998).

\bibitem{obradors88} X. Obradors,A. Labarta, A. Isalgue, J. Tejada, J. Rodriguez, and M. Pernet,
Solid State Commun. {\bf 65}, 189 (1988).

\bibitem{ramirez90} A. P. Ramirez, G. P. Espinosa, and A. S. Cooper,
Phys. Rev. Lett. {\bf 64}, 2070 (1990).

\bibitem{broholm90} C. Broholm,G. Aeppli, G.P. Espinosa, and A. S. Cooper,
Phys. Rev. Lett. {\bf 65},3173 (1990).

\bibitem{chalker92} J. T. Chalker, P.C.W. Holdsworth, and E. F.
Shender, Phys. Rev. Lett. {\bf 68}, 855 (1992).

\bibitem{schiffer95} P. Schiffer, A.P. Ramirez, D.A. Huse, P.L. gammel, U.Yaron, D.J. Bishop and A.J. Valentino. Phys. Rev. Lett.
{\bf 74}, 2379 (1995).

\bibitem{nakamura97} H. Nakamura,K. Yoshimoto, M. Shiga, M. Nishi, and K. Kakurai,
J Phys. Condens. Matter. {\bf 9}, 4701-4728 (1997).

\bibitem{canals00} B. Canals, and C. Lacroix, Phys. Rev. Lett {\bf
61}, 11251 (2000).

\bibitem{Zangrando1} M. Zangrando,M. Finazzi, G. Paolucci, B. Diviacco, R.P. Walker, D. Cocco, F. Parmigiani,
Rev. Scient. Instr. {\bf 72}, 1313 (2001).

\bibitem{galakov02} V. R. Galakhov,M. Demeter, S. Bartkowski, M. Neumann, N.A. Ovechkina, E.Z. Kurmaev,
N.I. Lobachevskaya, Y.M. Mukovskii, J. MiTchell, D.L. Ederer, Phys.
Rev. B {\bf 65}, 113102 (2002).

\bibitem{cowan81} R. D. Cowan,
The Theory of Atomic Structure and Spectra (University of California
press, Berkeley, 1981).

\bibitem{dirk} D. van der Marel and G.A. Sawatzky. Phys. Rev. B {\bf 37}, 10674 (1988)


\bibitem{zaanen} J. Zaanen and G.A. Sawatzky
Phys. Rev. B {\bf 33}, 8074 (1986).

\bibitem{kotani} H. Ogasawara, A. Kotani and K. Okada
Phys. Rev. B {\bf 43}, 854 (1991).

\bibitem{parmigiani} L. Sangaletti, F. Parmigiani and P.S. Bagus
Phys. Rev. B {\bf 66}, 115106 (2002).

\bibitem{gey} Gey-Hong Gweon, Je-Geun Park and S.-J. Oh
Phys. Rev. B {\bf 48}, 7825 (1993).

\bibitem{jarlborg} T. Jarlborg and P.O. Nilsson. J. Phys. C {\bf 12}, 265 (1979)

\bibitem{giappi} J.Y. Son, K. Okazaki, T.Mizokawa, A. Fuimori, T.
Kanomata and R. Note. J. Phys. Soc. Jap. {\bf 71}, 1728 (2002).

\bibitem{thole85} B. T. Thole, R.D. Cowan, G.A. Sawatzky, J. Fink, and J.C. Fuggle,
Phys. Rev. B {\bf 31}, 10 (1985).

\bibitem{thole91} G. van der Laan, and B. T. Thole,
Phys. Rev. B {\bf 43}, 13401 (1991).

\bibitem{thole92} B. T. Thole,P. Carra, F. Sette, G. van der Laan,
Phys. Rev. Lett. {\bf 68}, 1943 (1992).

\bibitem{carra93} P. Carra {\em et al.},B.T. Thole, M. Altarelli, X. Wang,
Phys. Rev. Lett. {\bf 70}, 694 (1993).

%


\end{thebibliography}
\end{document}